\documentclass[10pt,conference]{IEEEtran}

\usepackage{balance,subfigure,graphicx,epsfig,tabularx,multirow,amsfonts,epstopdf,float}
\usepackage{times}
\usepackage{hyperref}
\usepackage[ruled,vlined]{algorithm2e}
\usepackage{tikz}
\usetikzlibrary{arrows.meta}
\usepackage[utf8]{inputenc}
\usepackage{enumitem}
\usepackage{amsmath}
\usepackage{amssymb}
\usepackage{amsthm}
\usepackage{makecell} 

\newtheoremstyle{exampstyle}
  {0.5\topsep} 
  {0.5\topsep} 
  {} 
  {} 
  {\bfseries} 
  {.} 
  {.5em} 
  {} 

\theoremstyle{exampstyle}

\newcommand{\nop}[1]{}

\definecolor{xz}{rgb}{1,0,0}

\IEEEoverridecommandlockouts
\IEEEpubid{\makebox[\columnwidth]{978-1-6654-XXXX-X/25/\$31.00~\copyright~2025 IEEE \hfill} \hspace{\columnsep}\makebox[\columnwidth]{ }}

\begin{document}

\bstctlcite{IEEEexample:BSTcontrol}

\title{\textsc{ChronoConnect}: Tracking Pathways Along Highly Dynamic Vertices in Temporal Graphs}

\author{\IEEEauthorblockN{Jiacheng Ding, Cong Guo, Xiaofei Zhang}
\IEEEauthorblockA{University of Memphis, United States\\
\{jding2, cguo, xiaofei.zhang\}@memphis.edu}}

\maketitle

\begin{abstract}
With the proliferation of temporal graph data, there is a growing demand for analyzing information propagation patterns during graph evolution. Existing graph analysis systems, mostly based on static snapshots, struggle to effectively capture information flows along the temporal dimension. To address this challenge, we introduce \textsc{ChronoConnect}, a novel system that enables tracking temporal pathways in temporal graph, especially beneficial to downstream mining tasks, e.g., understanding what are the critical pathways in propagating information towards a specific group of vertices. 
Built on \textsc{ChronoConnect}, users can conveniently configure and execute a variety of temporal traversal algorithms to efficiently analyze information diffusion processes under time constraints. Moreover, \textsc{ChronoConnect} utilizes parallel processing to tackle the explosive size-growth of evolving graphs. We showcase the effectiveness and enhanced performance of \textsc{ChronoConnect} through the implementation of algorithms that track pathways along highly dynamic vertices in temporal graphs. Furthermore, we offer an interactive user interface for graph visualization and query result exploration. We envision \textsc{ChronoConnect} to become a powerful tool for users to examine how information spreads over a temporal graph.
\end{abstract}


\section{Introduction}

Temporal graphs~\cite{holme2012temporal} are emerging as a pivotal model in the realm of graph analysis, capturing the dynamic nature of real-world graph data and garnering significant research interest~\cite{DBLP:journals/tbd/ChengMJZWXW24, DBLP:journals/vldb/AghasadeghiBS24, DBLP:journals/pvldb/LayneCSG23}. These graphs are particularly instrumental in studying diffusion or information propagation~\cite{DBLP:conf/kdd/MyersZL12, Ciaperoni_2020}, where the dynamics of the network are crucial for understanding how information spreads over time. This has led to the adoption of temporal graphs in various applications, translating complex dynamic relationships into path queries to uncover patterns of information flow.

While existing literature has defined diverse path semantics to analyze temporal graphs~\cite{DBLP:journals/corr/HuangCW14,DBLP:journals/pvldb/WuCHKLX14}, the practical implementation of these theories into systems like ChronoGraph~\cite{DBLP:conf/data/HaeuslerTKFNB17}, Clock-G~\cite{DBLP:journals/tlsdkcs/Massri0PM23}, Auxo~\cite{DBLP:journals/bigdatama/HanLCC19}, Chronos~\cite{DBLP:conf/eurosys/HanMLWYZPCC14}, and TeGraph~\cite{DBLP:conf/icde/HuanLLLHCJWS22}, to name a few, marks a significant advancement in the field. These systems offer robust platforms for executing path queries over temporal graphs, yet there remains an overlooked aspect critical to understanding information diffusion: the role of highly dynamic vertices. In many scenarios, vertices that exhibit significant changes over time are key to the diffusion process, and prioritizing these vertices in path queries can yield insights that are not only more meaningful but also highly beneficial for downstream applications.

In this work, we introduce \textsc{ChronoConnect}, a prototype system designed to prioritize and incorporate these highly dynamic vertices in the evaluation of path queries. Recognizing that these vertices do not exhibit high degrees of change synchronously, \textsc{ChronoConnect} tackles the complex task of correlating these vertices across different timestamps, a challenge that cannot be addressed by mere extensions of existing models. To this end, we propose a methodology for extracting significant subgraphs that maintains the temporal correlation among dynamic vertices, enhancing our understanding of their role in information propagation.

Furthermore, \textsc{ChronoConnect} introduces innovative cross-snapshot path monitoring algorithms, allowing for a detailed examination of how information traverses through these dynamic vertices, especially towards specific target nodes, via an interactive visualized query output. This capability is crucial for observing and understanding the pathways of information spread in networks.
In addition to its specific focus on dynamic vertices, \textsc{ChronoConnect} represents a leap forward in temporal graph analysis by integrating temporal dimensions with graph traversal semantics. Moreover, by employing parallel processing, \textsc{ChronoConnect} harnesses modern multi-core hardware to enhance the data preprocessing and query performance significantly, addressing the limitations of traditional methods that rely on offline processing and snapshot analysis.

In essence, \textsc{ChronoConnect} not only addresses the critical need for efficient and meaningful analysis of information propagation in temporal networks but also enriches the toolkits of temporal graph analysis by offering a user-friendly and powerful tool for uncovering the nuanced patterns of dynamic networks.

\begin{figure}[t]
  \centering
  \includegraphics[width=\columnwidth]{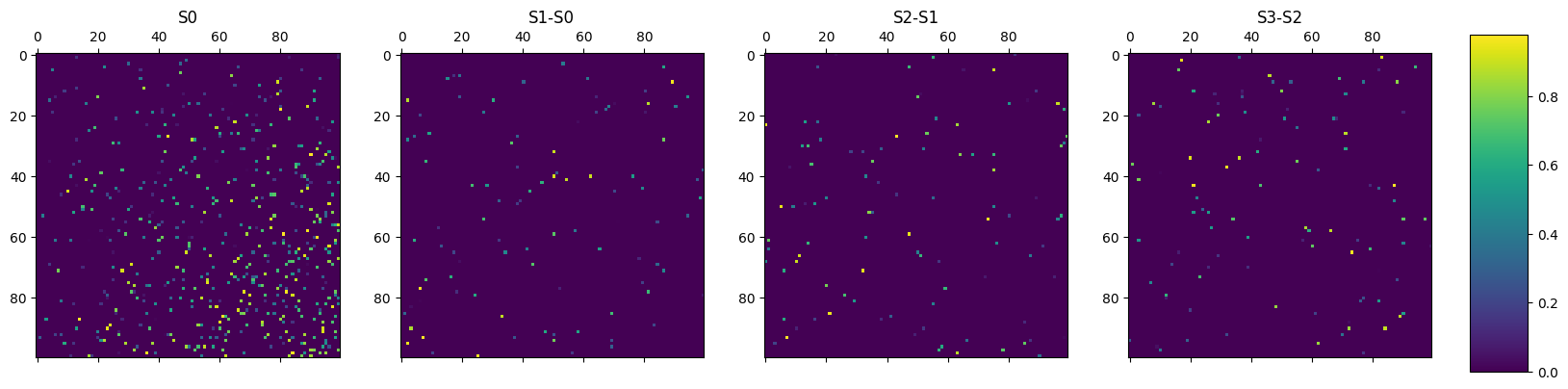}
  \caption{\small Illustration of a snapshot-based temporal graph emphasizing substructures with significant changes.}
  \label{fig:snapshot}
\end{figure}

\section{Solution Overview}

In this section, we first explain the data model and the major computation workload, then give an overview of \textsc{ChronoConnect}'s solution architecture and workflow.

\noindent\textbf{Preliminaries.} Adhering to the temporal graph model outlined in~\cite{holme2012temporal}, we analyze a temporal graph $G=\langle V,E,\mathcal{A}\rangle$ over time domain $T$, where each edge $e$ is tagged with timestamps $t_{start}$ and $t_{end}$. Here, $\mathcal{A}$ represents the comprehensive set of attributes for vertices and edges. For visualization and computational efficiency, we utilize a snapshot-based representation, transforming $G$ into sequential snapshots ${S_{t_0}, S_{t_1}, ..., S_{t_n}}$, ensuring that an edge $e$ exists in snapshot $S_{t_i}$ \textbf{iff} $e.t_{end} \leq t_i$. This method assumes edges in a snapshot have reached their destination.

We define highly dynamic vertices ($V_{HD}$) as those whose attributes or connectivity change markedly between consecutive snapshots. Leveraging established methods in bursting behavior detection~\cite{DBLP:conf/ijcai/ZhaoWYS019}, these vertices can be identified using existing algorithms or user-defined functions (UDFs) tailored to specific requirements. For illustration, Figure~\ref{fig:snapshot} displays the first graph snapshot and the highlighted $V_{HD}$ in following snapshots, aiding users in visualizing the graph's evolution.

Our study focuses on tracing information diffusion paths through the set of highly dynamic vertices, $V_{HD}$. These vertices may not be directly linked, as $V_{HD}$ can be dispersed across the graph, a scenario illustrated in Figure~\ref{fig:snapshot}. To address this, we create a \textit{significant subgraph} for each snapshot that includes $V_{HD}$, maintaining their connectivity to aid in path discovery. In \textsc{ChronoConnect}, these significant subgraphs are vital for capturing the key diffusion routes among $V_{HD}$.

We introduce an information diffusion path query, denoted as $Q=\langle q,V\rangle$, where $q$ is a query vertex from $V_{HD}$, and $V$ represents a set of target vertices for the diffusion process. Recognizing that diffusion dynamics can vary by application, our analysis adopts the shortest path framework for information traversal. We explore two types of query conditions to accommodate different application needs, detailed further in Section~\ref{sec:algorithms}.

\begin{figure}[t]
  \centering
  \includegraphics[width=0.9\columnwidth]{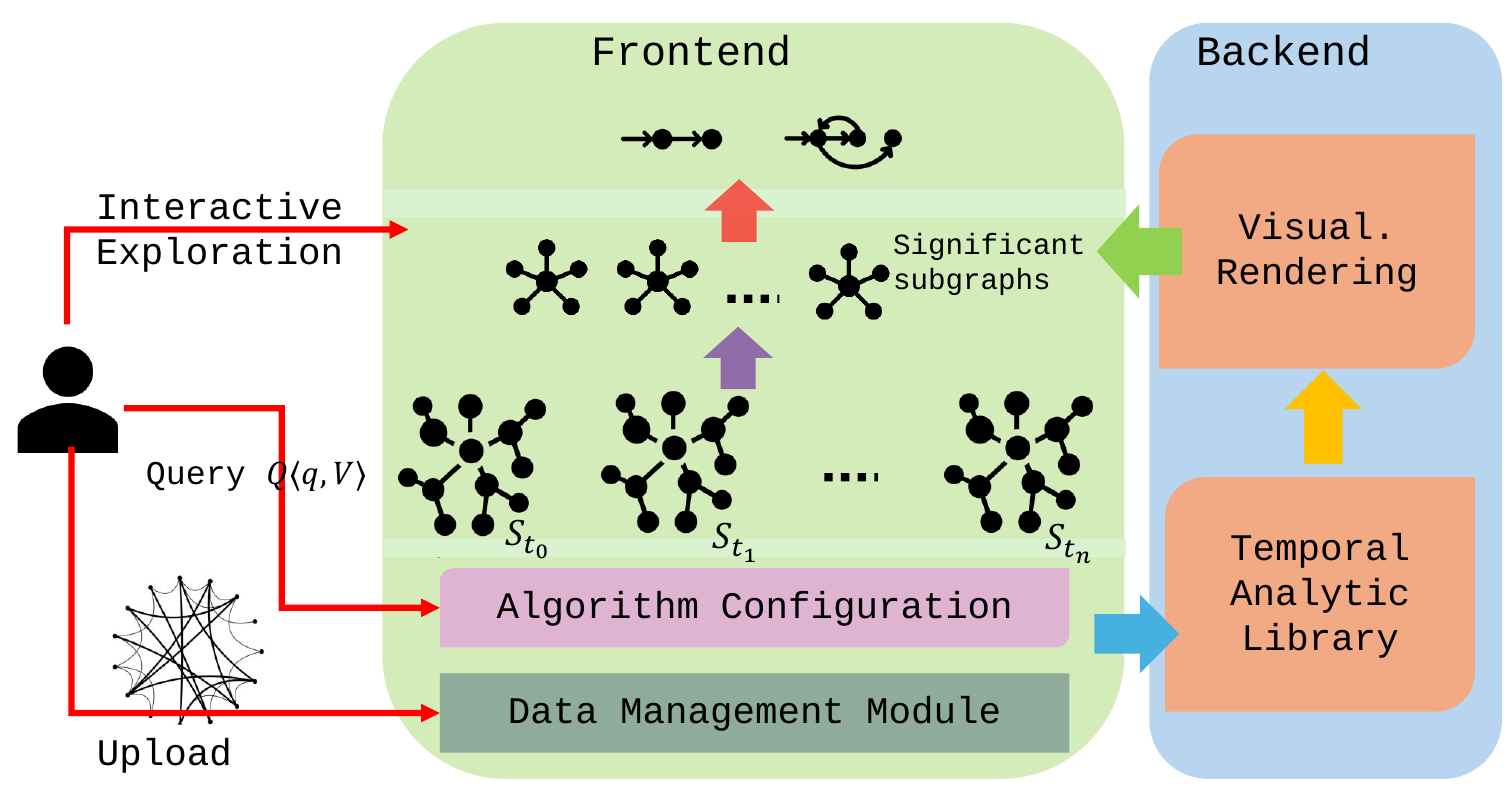}
  \caption{\small An overview of \textsc{ChronoConnect}.}
  \label{fig:framework}
\end{figure}

\label{subsec:architecture}
\noindent\textbf{Architecture and Workflow.} \textsc{ChronoConnect} integrates a streamlined frontend interface with a comprehensive backend analysis engine, as depicted in Figure~\ref{fig:framework}. Users interact with the frontend to upload temporal graph data, choose and set up temporal analysis algorithms, and visualize results. Conversely, the backend manages data preprocessing, algorithm execution, and result visualization. Here's a concise overview of \textsc{ChronoConnect}'s frontend and backend components:

\textsc{ChronoConnect} integrates three core modules into a cohesive frontend interface: the \textbf{Data Management Module} handles temporal graph ingestion and format transformation; the \textbf{Algorithm Configuration Module} provides an intuitive graphical interface for algorithm selection and parameter tuning; and the \textbf{Visual Analysis Module} renders interactive visualizations for exploring graph dynamics and temporal patterns. This unified design enables seamless workflow from data upload through analysis to insight discovery.

The backend analysis engine comprises two key components: the \textbf{Temporal Analytic Library}, which implements algorithms for vertex dynamicity detection, subgraph identification, and path pattern mining; and the \textbf{Visualization Rendering Module}, which transforms algorithmic outputs into interactive visual formats for intuitive exploration of temporal graph analytics.

In a typical use case, a user uploads a graph to the data management module, which then undergoes cleaning and indexing, transitioning to a snapshot-based format. Through the algorithm configuration module, users select algorithms to identify dynamic vertices and compute significant subgraphs, setting parameters like parallelism levels for efficiency.

The backend then processes the selected algorithms, analyzing snapshot collections to identify key structures and paths. The visualization module subsequently presents these findings, offering interactive exploration of dynamic vertices, connectivity, and diffusion paths. Real-time algorithm execution logs are provided to the frontend, allowing users to monitor and tweak their analyses. This iterative approach empowers users to engage deeply with the temporal graph data, enhancing their understanding and insights.

\section{Temporal Path Algorithms}
\label{sec:algorithms}

In this section, we elaborate on the key temporal graph analysis algorithms integrated into \textsc{ChronoConnect}. Initially, we describe the preprocessing phase, which filters out highly dynamic vertices $V_{HD}$ from each snapshot, followed by an in-depth examination of two distinct cross-snapshot shortest path algorithms that incorporate constraints from $V_{HD}$.

\subsection{Preprocessing}

\textsc{ChronoConnect}'s preprocessing takes two critical steps. The first step involves identifying vertices that undergo substantial changes across consecutive snapshots. To assess these changes, we enable a versatile, user-selectable metric configurable based on the study's needs. Options for this metric include Graphlet Count Distance (GCD), betweenness centrality, variations in edge weights, or metrics reflecting the structure of the ego network, such as local clustering co-efficiency. This metric should enable rapid computation over large graphs and support incremental updates to avoid full recalculations with each snapshot. In our current implementation, we choose a threshold-based approach based on the harmonic mean of the weights of incident edges on a vertex, aiming for a balance between computational efficiency and the ability to capture relevant changes.

It is worth noting that the vertices in $V_{HD}$ might not be directly connected in a single snapshot, as shown in Figure~\ref{fig:path_example}. As such, the second step of the preprocessing is to identify a subgraph that encapsulates all the vertices in $V_{HD}$, termed as the \textit{significant subgraph}. This process is vital for preserving the relationships among the vertices in $V_{HD}$. For this purpose, we employ the maximal k-core subgraph that captures $V_{HD}$ inclusively. Notably, the maximal k-core approach is well-suited for parallelization~\cite{DBLP:conf/icml/EsfandiariLM18}, enhancing \textsc{ChronoConnect}'s ability to efficiently process large graph datasets by leveraging this parallel computational capability.

\subsection{$V_{HD}$-aware Cross-Snapshot Shortest Path}

\textsc{ChronoConnect} aims to analyze the role of highly dynamic vertices ($V_{HD}$) across various timestamps in information propagation and their interrelations over time. We specifically explore the shortest path problem in temporal graphs, incorporating a $V_{HD}$ constraint to understand potential causal connections among $V_{HD}$ at different snapshots. This analysis is particularly relevant in scenarios like cyberattacks in the cybersecurity realm, where a sequence of compromised machines exhibits abnormal behavior over time. We focus on a shortest path problem tailored to paths traversing through $V_{HD}$ vertices, allowing us to track the evolving connectivity among these vertices. For practical application, we introduce a query, \texttt{FindChronopath}, $Q=\langle q,V\rangle$, where $q$ represents a source vertex and $V$ is a predefined set of target vertices in the graph. This query computes all $V_{HD}$-aware shortest paths from $q$ to $V$, offering insights into the dynamic flow of information from a specific source to multiple targets over time. Specifically, \texttt{FindChronopath} employs Dijkstra's algorithm on individual snapshots beginning from $q$ to ascertain the shortest routes to $V$, links these routes across successive snapshots to form potential paths, and then selects the shortest ones that meets the $V_{HD}$-aware constraint, designating them as critical paths for further analysis.

The \texttt{FindChronopath} query requires that each node on a critical path be a highly dynamic vertex, which may be too restrictive for certain contexts. In practical settings, incorporating some regular vertices into the path could enhance semantic consistency. Therefore, we propose a variant, \texttt{FindRelaxedChronopath}, $Q_{R}=\langle q,V\rangle$, which relaxes \texttt{FindChronopath}'s stringent condition by allowing intermediate nodes to include regular, non-highly dynamic vertices. This adjustment provides the algorithm with improved flexibility and better alignment with real-world semantic contexts. Additionally, \texttt{FindRelaxedChronopath} employs a significance scoring system to evaluate paths, favoring those with higher scores. This scoring considers the path length and the proportion of highly dynamic vertices, ensuring a balanced consideration of path length and vertices' dynamicity.

\begin{figure}[t]
    \centering
  \includegraphics[width=\columnwidth]{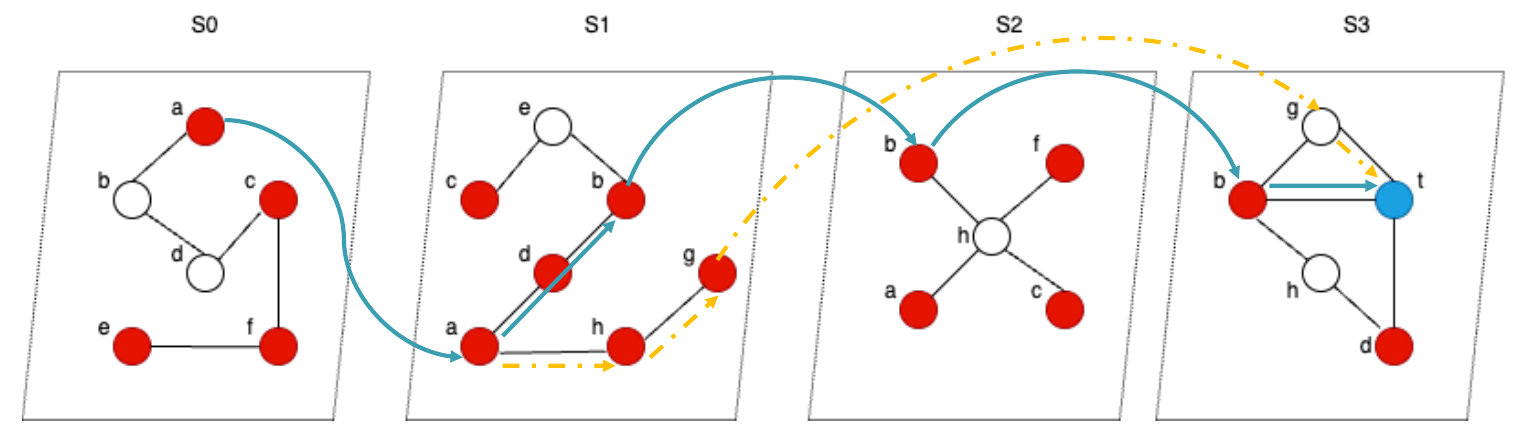}
  \caption{\small Illustration of chronopath across multiple snapshots. Red vertices are in $V_{HD}$. Non-highly dyanmic vertices (blank ones) may be needed to connect $V_{HD}$ in each snapshot.}
  \label{fig:path_example}
\end{figure}

Figure~\ref{fig:path_example} illustrates the results of $Q=\langle a,\{t\}\rangle$ and $Q_R=\langle a,\{t\}\rangle$. The path depicted by the solid blue line adheres strictly to the $V_{HD}$ constraint, with each node along this path being a highly dynamic vertex. Conversely, the dashed yellow line represents a path that bypasses snapshot $S_2$ and potentially offers a shorter route from $a$ to $t$. Here, vertex $g$ is not required to be a highly dynamic vertex across all snapshots to facilitate the information transfer from $a$ to $t$. This flexibility might align better with the semantic of various real-world scenarios.

\subsection{Summarizing $V_{HD}$-aware Propagation Paths}
Certain chronopaths may repeatedly appear within local regions, forming high-frequency propagation subgraphs. To reveal these local propagation patterns, we design the ExtractFrequentEdges algorithm to discover frequently occurring subpaths as a summarization of the $V_{HD}$-aware propagation paths.

ExtractFrequentEdges counts the frequency of each edge in all chronopaths found after executing $Q$. When an edge's frequency exceeds a given threshold, the algorithm extracts all paths containing that edge to form a propagation subgraph, reflecting high-frequency interaction patterns between the source vertex and the target vertices.

In summary, \textsc{ChronoConnect} offers a comprehensive set of temporal graph analysis algorithms that facilitate exploring key components and dynamic patterns of the underlying graph data from different perspectives. These modular algorithms can be applied independently or in combination to meet diverse analysis needs, laying the foundation for future extensions and optimizations.

\begin{figure*}[t]
    \centering
    \includegraphics[width=0.9\textwidth]{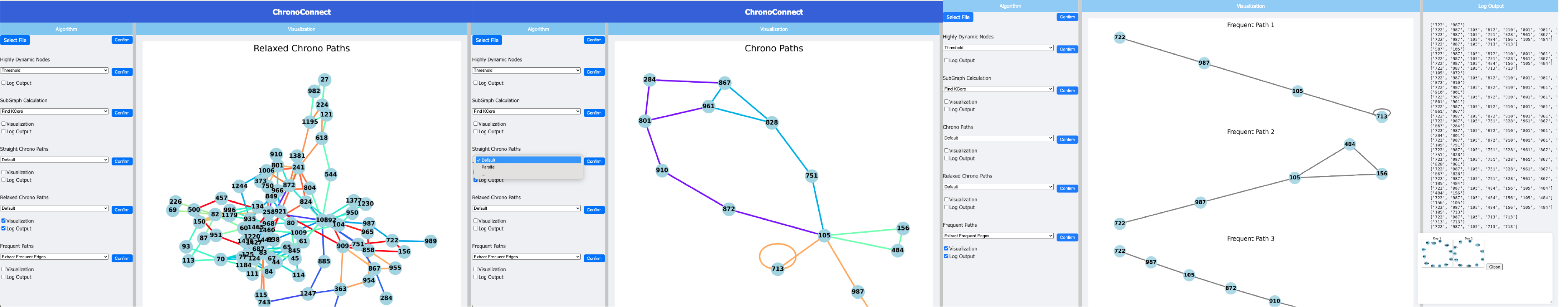}
    \caption{ChronoConnect demonstration interface showing: (a) Main dashboard with highly dynamic vertices ($V_{HD}$) identification and frequent path patterns visualization, including individual snapshot exploration feature; (b) FindChronopath query results displaying strict $V_{HD}$-constrained paths; (c) FindRelaxedChronopath query results showing flexible paths that may include non-dynamic intermediate vertices. The visualization aggregates paths sharing common vertices with varying edge colors to highlight different propagation patterns.}
    \label{fig:demo_combined}\vspace{-4ex}
\end{figure*}

\section{Demonstration}

\noindent\textbf{System Interface and Settings.} \textsc{ChronoConnect} provides an intuitive web interface for temporal graph analysis. Users can: (1) upload graph data meeting the specified format; (2) select and configure various graph processing algorithms; (3) visualize results and download query logs; (4) enable parallel processing for enhanced performance on multi-core systems.

Figure~\ref{fig:demo_combined} shows the complete demonstration interface, featuring the main dashboard for $V_{HD}$ identification, path query results visualization, and interactive exploration capabilities.

\noindent\textbf{Evaluation Datasets.} To validate \textsc{ChronoConnect}'s effectiveness, we evaluate it on three temporal networks from different domains:
\begin{itemize}
    \item Email-Eu-Core (EC)~\cite{Leskovec2005graphs}: A communication network with 986 nodes and 332,334 timestamped edges from a European research institution.
    \item Bitcoin OTC (BTC)~\cite{kumar2016edge}: A transaction network containing 5,881 nodes and 35,592 weighted, signed edges, representing cryptocurrency exchanges between anonymous users.
    \item Network Traffic (NT): An enterprise network flow dataset from our partner with 1,471 nodes and 5.5M timestamped edges, enabling scalability evaluation in cybersecurity scenarios.
\end{itemize}

The source code and a demonstration of ChronoConnect are available at \url{https://github.com/graphuofm/chronoconnect}.

\noindent\textbf{Case Study: Network Traffic Pattern Discovery.} We applied \textsc{ChronoConnect} to analyze the network traffic dataset spanning May-August 2023, creating daily snapshots to track machine interactions. As shown in Figure~\ref{fig:demo_combined}, our analysis revealed three predominant chain-like communication patterns among highly dynamic vertices, providing valuable insights for detecting potential malicious activities. 
The visualization aggregates chronopaths sharing common vertices (indicated by edge colors), revealing critical propagation patterns. This analysis successfully identified machines exhibiting abnormal communication behaviors across multiple time periods, demonstrating \textsc{ChronoConnect}'s practical utility in cybersecurity applications.

\noindent\textbf{Comparative Results.} Table~\ref{tab:combined_results} presents a quantitative comparison between \textsc{ChronoConnect} and traditional degree centrality-based approaches across all three datasets. For HDV detection, we set $w_1=0.8$, $w_2=0.2$, and $\theta=0.1$ with 10 temporal intervals. Traditional methods selected nodes with degree centrality $\tau>0.1$ as HDVs. Results demonstrate that \textsc{ChronoConnect} identifies significantly more dynamic vertices while achieving better coverage in most cases.  The longer average path lengths indicate our method captures more complex propagation patterns, essential for understanding information diffusion in temporal graphs. In the Bitcoin dataset, a few high-degree nodes dominate most transactions. Since \textsc{ChronoConnect} focuses on highly dynamic vertices, it yields lower coverage and shorter average path lengths. Thus, in networks like Bitcoin, where ordinary nodes show little dynamicity, \textsc{ChronoConnect} produces limited coverage and reduced path lengths.

\nop{
\begin{table}[t]
\vspace{-2ex}
\small
\centering
\caption{Performance Comparison Across Datasets}
\label{tab:combined_results}
\begin{tabular}{|l|c|c|c|}
\hline
\textbf{Method} & \textbf{HDV} & \textbf{Coverage} & \textbf{Avg Path} \\
 & \textbf{Count} & \textbf{Rate} & \textbf{Length} \\
\hline
\multicolumn{4}{|c|}{\textbf{Email-Eu-Core}} \\
\hline
\textsc{ChronoConnect} & 776 & 0.163 & 2.20 \\
Traditional & 5 & 0.080 & 1.00 \\
\hline
\multicolumn{4}{|c|}{\textbf{Bitcoin OTC}} \\
\hline
\textsc{ChronoConnect} & 10 & 0.084 & 0.80 \\
Traditional & 5 & 0.318 & 1.60 \\
\hline
\multicolumn{4}{|c|}{\textbf{Network Traffic}} \\
\hline
\textsc{ChronoConnect} & 709 & 0.205 & 2.45 \\
Traditional & 5 & 0.127 & 1.71 \\
\hline
\end{tabular}
\end{table}
}

\begin{table}[t]
\vspace{-2ex}
\small
\centering
\caption{Performance Comparison Across Datasets}
\label{tab:combined_results}
\begin{tabular}{|l|c|c|c|c|}
\hline
\textbf{Data} & \textbf{Method} & \makecell{\textbf{HDV} \\ \textbf{Count}} & \makecell{\textbf{Coverage} \\ \textbf{Rate}} & \makecell{\textbf{Avg Path} \\ \textbf{Length}} \\
\hline
\multirow{2}{*}{EC}  & \textsc{ChronoConnect} & 776 & \textbf{0.163} & \textbf{2.20} \\
                     & Traditional             &   5 & 0.080 & 1.00 \\
\hline
\multirow{2}{*}{BTC} & \textsc{ChronoConnect} &  10 & 0.084 & 0.80 \\
                     & Traditional             &   5 & \textbf{0.318} & \textbf{1.60} \\
\hline
\multirow{2}{*}{NT}  & \textsc{ChronoConnect} & 709 & \textbf{0.205} & \textbf{2.45} \\
                     & Traditional             &   5 & 0.127 & 1.71 \\
\hline
\end{tabular}
\end{table}
\vspace{-1ex}
\section*{Acknowledgment}
This work is partially funded by the NSF grant CCF-2217076. \vspace{-1ex}

\bibliographystyle{IEEEtran}
\bibliography{sample}

\end{document}